\documentclass[]{raa}
\usepackage{amssymb}            
\usepackage{graphicx,times}
\usepackage{natbib}

\providecommand{\bm}[1]{\mbox{\boldmath$#1$\unboldmath}}

\begin{document}

\title{The New RS CVn Binary V1034 Her Revisited and the orbital period - Activity relation of Short-period RS CVn binaries using photometric distortion amplitude
}

 \volnopage{ {\bf 2011} Vol.\ {\bf 9} No. {\bf XX}, 000--000}
   \setcounter{page}{1}

   \author{LiYun. Zhang
      \inst{1}
   }

   \institute {College of Science/Department of
Physics\&NAOC-GZU-Sponsored Center for Astronomy Research,\\ Guizhou
University, Guiyang 550025, P.R. China; {\it Liy\_zhang@hotmail.com}\\
\vs \no
   {\small Received 2011 Oct. 12; accepted 2011 Nov. 25 }
}

\abstract{This paper presents new CCD $BVRI$ light curves of the
newly discovered RS CVn eclipsing binary V1034 Her in 2009 and 2010,
which shapes are different from the previous published results. They
show asymmetric outside eclipse and we try to use a spot model to
explain the phenomena. Using the Wilson-Devinney program with
one-spot or two-spots model, photometric solutions of the system and
starspot parameters were derived. Comparing the two results, it
shows that the case of two spots is better successful in reproducing
the light-curve distortions. For all the spot longitudes, it
suggests that the trend towards active longitude belts and each
active longitude belts might be switch. Comparing the light curves
of 2009 and 2010, it indicates that the light curve changes on a
long time scale of one year, especially in phase 0.25. In addition,
we also collected the values of the maximum amplitudes of
photometric distortion of the short-period RS CVn binary. We found
for the first time that there is a trend of increasing activity with
decreasing the orbital period. Finally, fitting all available light
minimum times including our newly obtained ones with polynomial
function confirmed that the orbital
period of V1034 underwent up increase. \\
 }

   \authorrunning{Liyun Zhang }            
   \titlerunning{V1034 Her revisited and the orbital period - active relation }  
   \maketitle

%
%
\section{Introduction}           
\label{sect:intro}

\indent For V1034 Her (GSC 0983.1044, ROTSE 1 J165241.80+124905.2),
it is a newly identified RS CVn-type eclipsing binary, which is also
characterized by the light-curve asymmetries (Kaiser et al. 2002;
Do$\check{g}$ru et al. 2009; Ordway \& Van Hamme 2004). Therefore,
it is a very intriguing object to study stellar magnetic activity
and test the effect of magnetic braking. However, it is poorly
studied by
astronomers, so we re-observed it in two different seasons to discuss starspot activity. \\
\indent V1034 Her was discovered by Akerlof et al. (2000) as a 13th
mag eclipsing binary with a period of 0.40763 day and an amplitude
of 1.022 mag from the ROTSE 1 sky survey. Later, Kaiser et al.
(2002) revised the standard magnitude $V=12.90$, and color indices
$B-V~=~0.78$. They also derived the linear ephemeris of the system
with the period of 0.8153 day and suggested that V1034 Her is a RS
CVn type eclipsing detached binary due to the light-curves
variations, which were explained using two-spots model on the
secondary component. Later, Ordway and Van Hamme (2004) also
obtained the $BVRI$ light curves of the system and analyzed them
using Wilson-Devinney program with the spot model. Recently,
Do$\check{g}$ru et al. (2009) revised the ephemeris and considered
firstly the possibility orbital period variation. The O-C diagram
shows an upward parabola, which indicates a secular increase in the
orbital period of the system. They also confirmed it the characters
of RS CVn type and explained them in terms of two
large dark starspots on the primary component.\\
\indent As our on-going program of the study of RS CVn binary with
multi-color photometry and high-resolution spectra (Gu et al. 2002,
Zhang \& Gu 2007, Zhang \& Gu 2008, Zhang et al. 2010a, etc), we
present new $B, V, R,$ and $I$ CCD light curves of V1034 Her in two
different seasons and analyze them to study the properties of active
regions using the version 2003 of Wilson-Devinney code. In addition,
we accumulated the maximum amplitudes of photometric distortions of
short-period RS CVn binary from the literatures to discuss the
relation between the photospheric starspot activity and
the orbital period.\\

\section{Observations}
\label{sect:Obs} The new photometric observations were made on seven
nights (2009 March 22, 24, 25, 26, 30, and 31, and April 1) and two
nights ( 2010 March 26 and 27) with the 85-cm telescope of Xinglong
station of the National Astronomical Observatories of China. The
camera was equipped with a $1024 \times 1024$ pixel CCD and the
standard Johnson-Cousin-Bessell $BVRI$ filters (Zhou et al. 2009).
Our observations of V1034 Her was made in $B$, $V$, $R$, and $I$
bands. All observed CCD images were reduced using the Apphot
sub-package of IRAF in the standard fashion (including image
trimming, bias subtraction, flat-field division, cosmic ray removal,
and aperture photometry). The comparison star GSC0983-566 and the
check star ($\alpha_{2000}$=$16^h$$52^m$$19^s$.07;
$\delta_{2000}$=$12\,^{\circ}$$50'$$43''$.9) were chosen near
target. The LCs of V1034 Her are plotted in Figure.~1, where
circles($\circ$) indicate Mar. 26 - Apr. 1 2009, squares($ \Box $) -
Mar. 26-27 2010. The errors of individual points are better than
0.01 mag in all bands.\\
\indent From our observations, two primary minimum times
(2454916.3181$\pm$0.0044; 2454921.2101$\pm$ 0.0010) and one
secondary minima (2454923.2477$\pm$ 0.0003) were obtained using the
method of Kwee and van Woerden (1956). To calculate an update
ephemeris and period variation, all published minimum times were
collected from the literatures (Table.1). Using the CCD minima
times, an updated linear ephemeris formula was obtained as follows:
\begin{eqnarray}
Min.I~=~HJD 2451767.6630(\pm 0.0003)+0^{d}.8152912(\pm 0.0000001)
days E.
\end{eqnarray}
For our new observations, the phases of data points are calculated
using the above equation~(1).\\
\indent Do$\check{g}$ru et al. (2009) found the (O-C) (observational
times of light minimum calculational times of light minimum) diagram
shows an upward parabola. To revise the period change, we reanalyzed
them with our new minimum. During the analysis, the weights 0, 1 and
5 were given to visual, photographic and CCD observations. The two
CCD minimum times (deviating too much) whose $O-C$ values are higher
than 0.0070 were omitted because they might be influenced by spots
and shifted. They are listed under the line at the end of Table.1.
Then, with the least-square method, the following quadratic
ephemeris was obtained.
\begin{eqnarray}
Min.I~=~JD(Hel.)2451767.6632(\pm 0.00008)+0^{d}.81529075(\pm
0.00000004)E
\nonumber\\
+1^{d}.38(\pm 0.08)\times10^{-10}\times E^{2}.
\end{eqnarray}
It indicates that the orbital period shows a continuous secular
increase of the period at a rate of dp/dt=$1.24(\pm 0.07)\times
10^{-7}dyr^{-1}$ (or $0.0107(\pm 0.0006)$ second per year).
The $O-C$ values are listed in the third column of Table 1 and plotted in Figure 2. \\

\begin{table}
\caption{The minimum times of V1034 Her.}
\begin{tabular}{lrrll}
\hline \hline \multicolumn{1}{l}{JD(Hel.)} &
\multicolumn{1}{c}{Cycle} & \multicolumn{1}{c}{(O-C)}  &
\multicolumn{1}{c}{Method}
&\multicolumn{1}{c}{Reference} \\
\hline
2445901.652&$-7195.0$&$-0.0043$&pg&Kaiser et al. (2002)\\
2446116.891&$-6931.0$&$-0.0066$&pg&Kaiser et al. (2002)\\
2451738.7197&$-35.5$&$-0.0016$&CCD &Kaiser et al. (2002)\\
2451740.7594&$-33.0$&$-0.0001$&CCD &Kaiser et al. (2002)\\
2451747.6883&$24.5$&$-0.0012$&CCD &Kaiser et al. (2002)\\
2451753.8028&$-17.0$&$-0.0014$&CCD &Kaiser et al. (2002)\\
2451767.6641&$0.0$&0.0002&CCD &Kaiser et al. (2002)\\
2452084.8118&$389.0$&$0.0004$&CCD &Kaiser et al. (2002)\\
2452873.1962&$1356.0$&$-0.0021$&CCD &Krajci (2005)\\
2453130.4213&$1671.5$&$0.0012$&CCD &Krajci (2005)\\
2453518.5010     & $2147.5$ & $0.0001$ & CCD &H$\ddot{u}$bscher et al. (2006)\\
2453596.3613    & $2243.0$ &$0.0002$   & CCD &Bakis et al. (2005) \\
2453872.3376    & $2581.5$ &$0.0006$   & CCD &Do$\check{g}$ru (2009)  \\
2453874.3761   & $2584.0$ &$0.0008$   & CCD &Do$\check{g}$ru (2009)  \\
2453922.4776    & $2643.0$ &$0.0002$   & CCD &Do$\check{g}$ru (2009)  \\
2454200.4918    & $2984.0$ &$0.0003$   & CCD &H$\ddot{u}$bscher (2007)  \\
2454916.3181&3862.0&$0.0013$&CCD &present paper\\
2454921.2101&3868.0&$0.0016$&CCD &present paper\\
2454923.2477&3870.5&$0.0010$&CCD &present paper\\
\hline
2451265.8420  &    $-615.5000 $  &  $-0.0112$  & CCD&Diethelm (2001)\\
2451311.9100  &   $ -559.0000$  &   $-0.0071$  &  CCD&Diethelm (2001)\\
2452792.489&$1257.0$&$-0.0043$&vis&Diethelm (2003)\\
2453149.585     &$1695.0$ &$-0.0031$   &vis&Diethelm (2004)\\
2453440.647    &$2052.0$ &$0.0064$  &vis&Locher (2005)\\
\hline
\end{tabular}
\end{table}

   \begin{figure}[h!!!]
   \centering
   \includegraphics[width=8.0cm, angle=0]{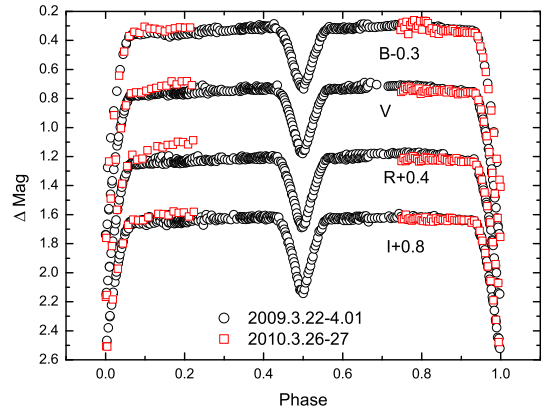}

   \caption{B, V, R, and I observations of V1034 Her in 2009 and 2010 at Xinglong
station. Circles($\circ$) indicate Mar. 22 - Apr. 1 2009, squares($
\Box $ ) - Mar.26 -27 2010. }
   \label{Fig1}
   \end{figure}

   \begin{figure}
     \begin{center}
    \includegraphics[angle=0,scale=0.8]{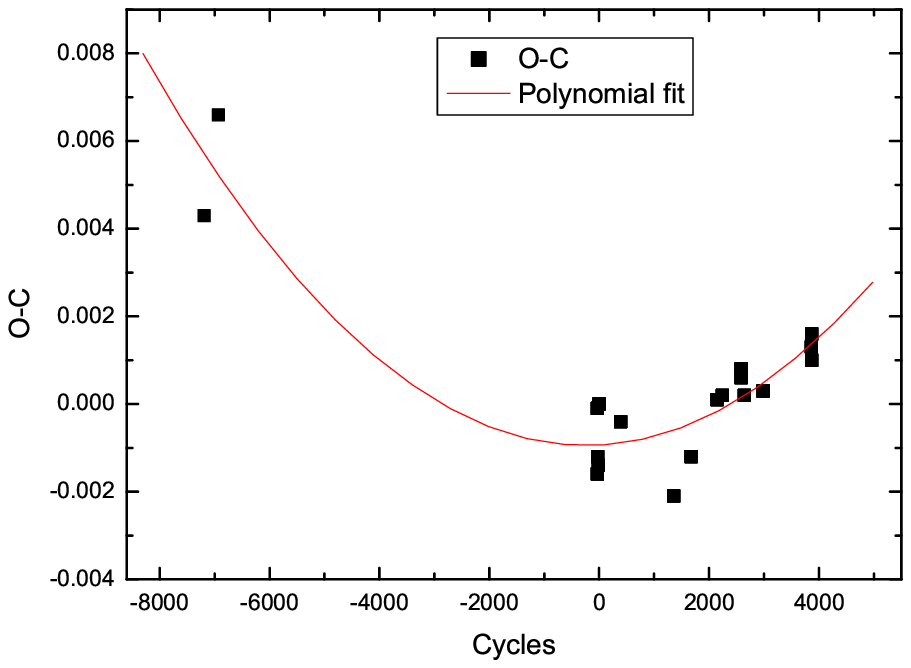}
   \caption{(O-C) diagram for the minimum
times of V1034 Her. The solid line represents the quadratic fitting
and show a period increase. \label{fig:2}}
 \end{center}
   \end{figure}

\section{Photometric analysis of V1034 Her}
\label{sect:data}

Because our data have high time-resolution and full phase coverage
in 2009, we might re-obtained the photometric solution of V1034 Her
using the 2003 version of the Wilson-Devinney program (Wilson \&
Devinney 1971, Wilson \& Devinney 1979, Wilson 1990 and 1994, Wilson
\& Van Hamme 2004, etc).\\
\indent Due to the asymmetric light curves outside the primary
eclipse, there is at least a large distortion in the phase range
0.1-0.3 (spot1). From a comparison between the observational light
curves and the clean theoretical ones (the system parameters with
less distortional effect are taken from the papers of Kaiser et al.
(2002) and Do$\check{g}$ru et al. (2009)), it indicates that there
might be a depression around 0.9 (spot2). Hence, we are not able to
tell whether there are one spot or two spots. Moreover, because of
having no spectroscopic observation, it is very difficult to tell
whether the spots are located on the primary or secondary
components, especially spot on the quadratures. Therefore, we
performed two solutions (one is one-spot model on the primary, the
other is two-spots model on the primary.) to search the final
result. The details of the procedure of photometric solution are
similar to those of our previous works of RT And (Zhang \& Gu 2007),
DV Psc (Zhang et al. 2010a),
GSC 3576-0170(Zhang et al. 2010b) and KQ Gem (Zhang 2010).\\
\indent All BVRI light curves are analyzed simultaneously. In
computing the photometric solutions, mode 2 of Wilson-Devinney code
(appropriate for detached binaries) is employed with the synchronous
rotation and zero eccentricity. Simple treatment is used to compute
the reflect effect, and the linear limb-darkening law is used to
compute the limb-darkening effect. The bolometric albedo
$A_{1}$=$A_{2}$=0.5 (Rucindki 1973), the limb-darkening coefficients
$x_{1B}$~=~0.819 $x_{2B}$~=~0.906, $x_{1V}$~=~0.686
$x_{2V}$~=~0.763,$x_{1R}$~=~0.568 $x_{2I}$~=~0.629, $x_{1I}$~=~0.467
$x_{2I}$~=~0.512 from Van Hamme (1993) and the gravity-darkening
coefficients $g_{1}$~=~$g_{2}$~=~0.32 for convective envelops (Lucy
1967) are set for the primary and the secondary, as usual. According
to the color index of $B-V$~=~0.78 obtained by Kaiser et al. (2002),
the effective temperature of the primary is $T_{1}$~=~5360~K from
Budding \& Demircan (2007)
(see Do$\check{g}$ru et al. 2009).\\

\indent We have performed two solutions: one-spot model and
two-spots model on the primary. The preliminary values of the
orbital parameters are taken from the prior photometric solutions
derived by Kaiser et al. (2002) and Do$\check{g}$ru et al. (2009).
We separately adjusted the orbital parameters and spot parameters
until the theoretical curves fit the observed ones well. For the
two-spots model, we fixed the spot latitude and temperature to avoid
correlation among the spot latitude, temperature and radius.
According to the temperature relation of the starspot and its
located photosphere for active stars (Berdyugina 2005), we could
determined that the starspot temperature is about 1500~K below the
photosphere of the primary for V1034 Her. Therefore, the temperature
factor is assumed to be 0.75. The preliminary latitude was set at
the intermediate latitude $50 ^{\circ}$ based on the prior result
for the RS CVn binary. After a lot of runs, two photometric
solutions of V1034 Her are derived and listed in Table~2. While, the
weighted sum of squares of residuals of two-spots model is smaller
than that of one-spot model, so we concluded that the two-spots
model being on the primary is better successful for describing the
LCs of V1034 Her. The theoretical light curves for the spotted
solutions and the observed light curves are shown in Figure.~3.
Corresponding
configurations for V1034 Her in phases 0.0, 0.25, 0.5 and 0.75 are shown in Figure.~4.\\
\begin{table}
\caption{Photometric solutions for V1034 Her.}
\begin{center}
\small
\begin{tabular}{cccc}
\hline \ Parameters & Values (one spot) & Values (two
spots)\\\hline\hline
$T_{1}$(K) & 5360$^a$& 5360$^a$\\
$T_{2}$(K) & 4823$\pm$5 & 4857$\pm$4\\
$i(deg)$     & 87.57$\pm$0.08 & 88.50$\pm$0.09 \\
$q$       & 1.006$\pm$0.007 & 0.98$\pm$0.01 \\
$\Omega_{1}$ & 5.825$\pm$0.040    &5.560$\pm$0.024 \\
$\Omega_{2}$ & 5.681$\pm$0.042    & 5.583$\pm$0.047 \\
$L_{1}/(L_{1}+L_{2})$ (B) &0.6688$\pm$0.0017&0.7063$\pm$0.0009\\
$L_{1}/(L_{1}+L_{2})$ (V) &0.6291$\pm$0.0020&0.6716$\pm$0.0011\\
$L_{1}/(L_{1}+L_{2})$ (R) &0.6005$\pm$0.0022&0.6462$\pm$0.0012\\
$L_{1}/(L_{1}+L_{2})$ (I) &0.5795$\pm$0.0023&0.6275$\pm$0.0013\\
$r_{1}(pole)$&0.2066$\pm$0.0017 & 0.2174$\pm$0.0012\\
$r_{1}(point)$&0.2120$\pm$0.0019 & 0.2240$\pm$0.0014\\
$r_{1}(side)$&0.2085$\pm$0.0018 & 0.2197$\pm$0.0013\\
$r_{1}(back)$&0.2111$\pm$0.0019 & 0.2228$\pm$0.0014 \\
$r_{2}(pole)$&0.2138$\pm$0.0019   & 0.2015$\pm$0.0020\\
$r_{2}(point)$&0.2200$\pm$0.0022 & 0.2064$\pm$0.0022\\
$r_{2}(side)$&0.2159$\pm$0.0020   & 0.2031$\pm$0.0020\\
$r_{2}(back)$&0.2189$\pm$0.0021   & 0.2055$\pm$0.0021\\
$latitude _{spot1}(deg)$&50.0$\pm$12.0 & 50$^a$\\
$longitude _{spot1}(deg)$&68.1$\pm$5.1& 70.2$\pm$5.1\\
$radius _{spot1}(deg)$&14.6$\pm$5.1& 15.3$\pm$0.4\\
$temperature f_{spot1}$&0.82$\pm$0.09& 0.75$^a$\\
$latitude _{spot2}(deg)$&- & 50$^a$\\
$longitude _{spot2}(deg)$&-& 333.9$\pm$2.6\\
$radius _{spot2}(deg)$&-& 10.9$\pm$1.6\\
$temperature f_{spot2}$&- & 0.75$^a$\\
\hline
$\sum_{i}$$(O-C)_{i}^{2}$&0.2862& 0.2238\\
\hline
\end{tabular}

Parameters not adjusted in the solution are denoted by a mark
``$^a$". \end{center}
\end{table}
\begin{figure}
  \begin{center}
    \includegraphics[angle=0,scale=0.8]{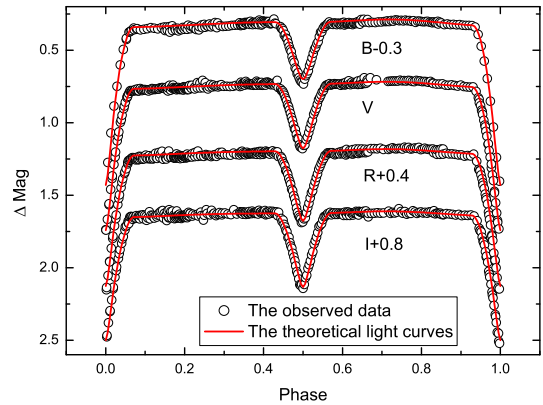}
   \end{center}
    \caption{The observational and theoretical light curves of V1034 Her in 2009. The circles and solid lines represent the observational and theoretical
    light curves, respectively.}\label{fig:3}
\end{figure}
\indent As can be seen from Fig.~1, the LCs do not have full phase
coverage in 2010 due to weather condition. Comparing of the LCs of
2009 and 2010, it indicates a small variations around 0.2 and 0.8 on
a long time scale of one year (see fig.1). The variations might be
explained by the spot variations. However, we could not derived the
spot
parameters because one small spot could affect more than half of the LC, and our light curves have very large data gaps in 2010.\\
\begin{figure}
  \begin{center}
    \includegraphics[width=10.5cm,height=7cm]{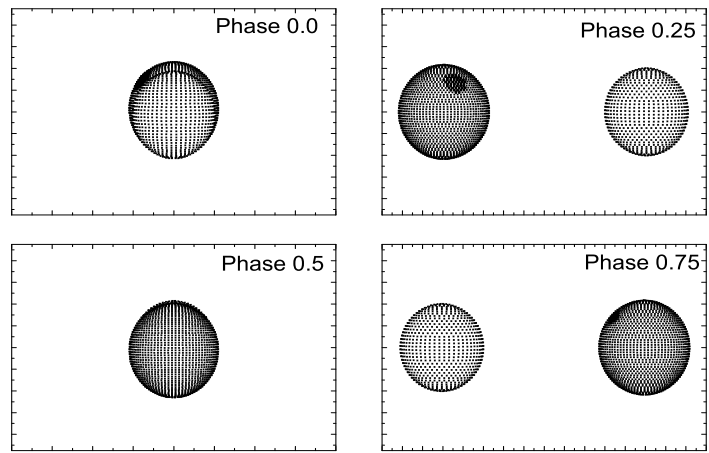}
    \begin{scriptsize}
\caption{The configurations of V1034 Her in phases 0.0, 0.25, 0.5
and 0.75.}\label{fig:4}
\end{scriptsize}
   \end{center}
\end{figure}

\section{Discussion and Conclusion}
In this paper, our new light curves in BVRI bands were analyzed
using the 2003 version of the Wilson-Devinney code with a spot
model. New absolute physical parameters and starspot parameters were
obtained. Secondly, we discussed the relation between photospheric
starspot activity and the orbital period for the short-period RS CVn
binary. Finally, the variation of the orbital period of V1034 Her were reanalyzed.\\
\subsection{Photometric solution and starspot activity of V1034 Her}
\indent The marked asymmetry of the light curves of V1034 Her might
suggest high-level surface activity, which was explained by
starspot. From our results, we derived that the case of two-spots
model being on the primary star is better successful in representing
the distortions of the light curves. For the orbital parameters, the
contribution of the primary component of V1034 Her to the total
light is 0.71 in $B$, 0.67 in $V$, 0.65 in $R$ and 0.63 in $I$ band.
Our new orbital parameters of luminosity ratios, the temperature of
the secondary, and the orbital inclination are similar to those
derived by Kaiser et al. (2002) and Do$\check{g}$ru et al. (2009).
However, the dimensionless potentials of the primary and secondary
components $\Omega _{1}$ and $\Omega _{2}$ are a bit smaller than
the results derived by Kaiser et al. (2002) and Do$\check{g}$ru et al. (2009). These may be affected by the starspot. \\
\indent It is well known that active-region longitude is the most
reliable spot parameters determined by the traditional Light-curve
method. For the result derived by Kaiser et al. (2002), the
longitudes of two spots in 2001 are about $118^{\circ}$ and
$335^{\circ}$, respectively. The spot longitude was transformed to
the binary orbital motion system. Later, Do$\check{g}$ru et al.
(2009) derived that the longitudes of the two cool spots were
$135^{\circ}$ and $239^{\circ}$ in 2006, respectively. For our
starspot parameters, the longitude of the spot1 is $70^{\circ}$ and
the spot2 is about $334^{\circ}$. Therefore, it suggested that
active regions tend to appear in two active longitude belts
$90^{\circ}$ and $270^{\circ}$, and each active longitude might be
migrate in the orbital reference frame (see Fig.5). The active
longitude phenomenon have also been found using photometry on many
other active RS CVn binary systems, such as EI Eri (Berdyugina \&
Tuominen 1998), $\sigma$ Gem (Henry et al. 1995), HK Lac (Olah Hall
\& Henry 1991), DV Psc (Zhang et al. 2010a) SV Cam (Zeilik et al.
1988) and RT And (Pribulla et al. 2000, Zeilik et al. 1989,
Zhang \& Gu 2007).\\
\begin{figure}
  \begin{center}
    \includegraphics[angle=0,scale=1.0]{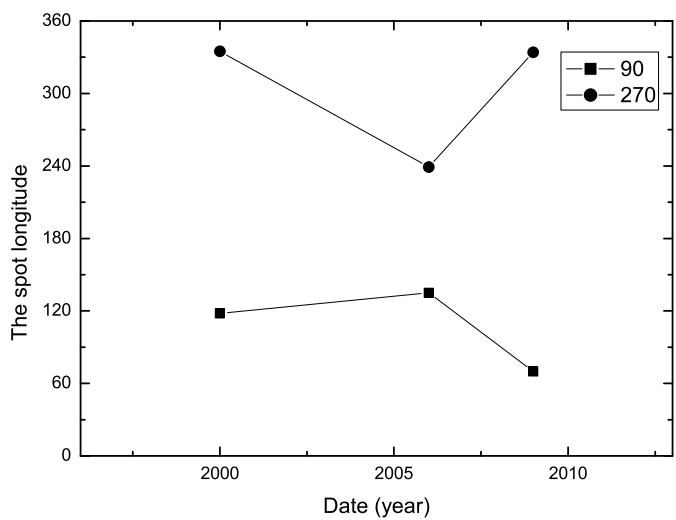}
    \begin{scriptsize}
\caption{The spot longitudes from 2000 to 2009.}\label{fig:5}
\end{scriptsize}
   \end{center}
\end{figure}
\indent Comparasion of the LCs of 2009 and 2010, it indicates the
spot1 becomes weaker at phase 0.2 and the spot2 becomes a bit
stronger at phase 0.8 in 2010. Therefore, the spot activity changed
on long time scale of one year. Indeed, Kaiser et al. (2002) also
found the light curves differ on the order of 0.04 magnitudes at
phase 0.9 during their observations in 2000 and 2001. These
indicates starspot variation on
a time scale of one year.\\
\subsection{The orbital period - Activity relation for the short period RS CVn binary}
For our observational light curves of V1034 Her, Max.II is around
0.04 mag in $B$, 0.07 mag in $V$, 0.06 mag in $R$ and 0.04 mag in
the $I$ band brighter than Max.I, respectively. As you know, the
light curve distortions are caused by dark photospheric spots for RS
CVn stars. In order to detect the relation of the orbital period and
the photospheric activity, we have also collected the values of the
maximum amplitude of photometric distortions of other short-period
RS CVn binary(Strassmeier et al. 1993; Eker et al. 2008). The values
are listed in Table 3, which includes the object name, spectral
type, the orbital period, the maximum distortion amplitudes in BVRI
bands. For these values, the number of BV bands values are more than
the RI bands. To detect the amplitude variation with the orbital
period, they are plotted in Figure.6. As can be seen from Fig.6,
there is a trend of increasing activity with decreasing the orbital
period, and the trend are basically consistent with each other in
the B and V bands (see Fig.6). To clarify the change, we also used
the linear least-squares method to fit, and the linear formulas were
obtained for B and V bands as follows:
\begin{eqnarray}
The ~ B~ amplitude(mag)~=0.257(\pm 0.067) \nonumber -0.223(\pm0.106)
\times P.\\
The ~ V ~ amplitude(mag)~=0.151(\pm 0.052) \nonumber -0.068(\pm
0.083)\times P.\\
\end{eqnarray}
These fits are also plotted in figure~6 with solid line. They are
consistent with the rotation-activity correlation of RS CVn binaries
by using chromospheric activity indicators (Montes et al. 1995; Montes et al. 1996),
the radio luminosity (Gunn 1998), the extreme ultraviolet emission (Mitrou et al. 1997).\\

\subsection{The period variation of V1034 Her}
\indent The variation of the orbital period of V1034 was reanalyzed
on the basis of our new minimum times and those collected from the
literatures. The quadratic term indicates that the orbital period of
V1034 Her shows a continuous increase at a rate of $0.0107(\pm
0.0006) second yr^{-1}$. The rate is similar to the result derived
by Do$\check{g}$ru et al. (2009). Since V1034 Her is a detached
system, it is impossible that the mass transfer directly from the
less massive component to the larger massive primary. So it is
probable that the coronal mass flow from the less massive component
to the large massive component by stellar wind (G$\acute{a}$lvez et
al., 2007). Because the photometric mass ratio is about 1, so it is
difficult to tell the massive component is the primary star or the
secondary component. Because the time range of (O-C) information is
rather short - about 25 years, it is too early to decide about the
character of the period variation. The upward parabola maybe
a part of long periodic oscillation cause by a presumed third component or by magnetic activity cycle (Applegate 1992; Lanza et al. 1998). More new observations are needed to confirm it in the next 20-30 year. \\
\begin{table}
\scriptsize\tabcolsep 0.20truecm \caption{The maximum distortion
amplitudes of short-period RS CVn binary in B, V, R and I bands.}
\begin{tabular}{llllllll}
\hline \hline \multicolumn{1}{l}{Object} &
\multicolumn{1}{c}{Spectral type} & \multicolumn{1}{c}{Period}  &
\multicolumn{4}{c}{The distortion amplitude (mag)}&\multicolumn{1}{c}{Reference} \\
\cline{4-7} &&(days)&B&V&R&I\\

\hline
RT And                     &F8-G0V+K1-3V&0.629    &0.095    & 0.075   &       &  0.05    & Pribulla et al., 2000; Zhang \& Gu 2007                                                                 \\
V1034 Her                  &G5-9V+K1-4V &0.815    &0.040    & 0.070   & 0.06  &  0.04    & Kaiser et al., 2002; Do$\check{g}$ru et al.,2009; my paper                                                           \\
DV Psc                     &K4V+        &0.309    &0.122    & 0.109   & 0.09  &  0.06    & Zhang \& Zhang 2007; Zhang et al., 2010a                                          \\
XY Uma                     &G2-5V+K5V   &0.479    &0.292    & 0.251   & 0.12  &  0.10    & Pribulla et al., 2001; Yuan 2010                                                       \\
CG Cyg                     &G0V+K3V     &0.631    &0.170    & 0.140   & 0.05  &  0.05    & Heckert 1996; Af\c{s}ar et al., 2004; Kozhevnikova et al., 2007a                          \\
BH Vir                     &F8V+G5V-M2V &0.817    &0.030    & 0.100   &       &          & Xiang et al., 2000; Kozhevnikova et al., 2007b      \\
WY Cnc                     &G5V+K7V    &0.830    &         & 0.110   &       &          & Kozhevnikova et al., 2007b                                                      \\
GSC03377-0296              &K3+         &0.422    &0.140    & 0.120   & 0.10  &          & Lloyd et al., 2007                                                              \\
DK CVn                     &K7V+M       &0.495    &0.159a   & 0.117a  & 0.10a &  0.08a   & Koff et al., 2005; Terrell et al., 2005                                                           \\
SV Cam                     &G0-G5V+K4V  &0.593    &0.150a   & 0.150   &       &          & Pribulla et al., 2003                                                          \\
UCAC3 295-68871            &            &0.461    &         &         & 0.10a &          & Solovyov et al., 2011                                                          \\
GSC 2038.0293              &G6-9+K1-3   &0.495    &0.090    & 0.070   & 0.06  &          & Bernhard \& Frank 2006                                                           \\
UV Psc                     &G5V-8+K2-3V &0.861    &0.094    & 0.100   &       &          & Akan et al., 1988; Kjurkchieva et al., 2005                                    \\
ER Vul                     &G0V+G5V     &0.698    &0.080a   & 0.050   &       &          & Heckert 1991; Qian 2001; Pribulla et al., 2003                                                                   \\
BB Scl                     &K3-5V+K4V   &0.477    &         & 0.070   &       &          & Bromage \& Buchley 1996; Watson et al., 2001                                                            \\
UV Leo                     &G0V+G2V     &0.600    &         & 0.050   &       &          & Kjurkchieva \& Marchev 2007                                                    \\
BC Phe                     &G5-8V+K3-5V &0.649    &         & 0.170   &       &          & Cutispoto 1995                                                                 \\

\hline
\end{tabular}
The values are denoted by a mark ``a", which are approximately
calculated by the light curves of the literatures.
\end{table}

\begin{figure*}
\begin{center}
\includegraphics[angle=0,scale=0.405]{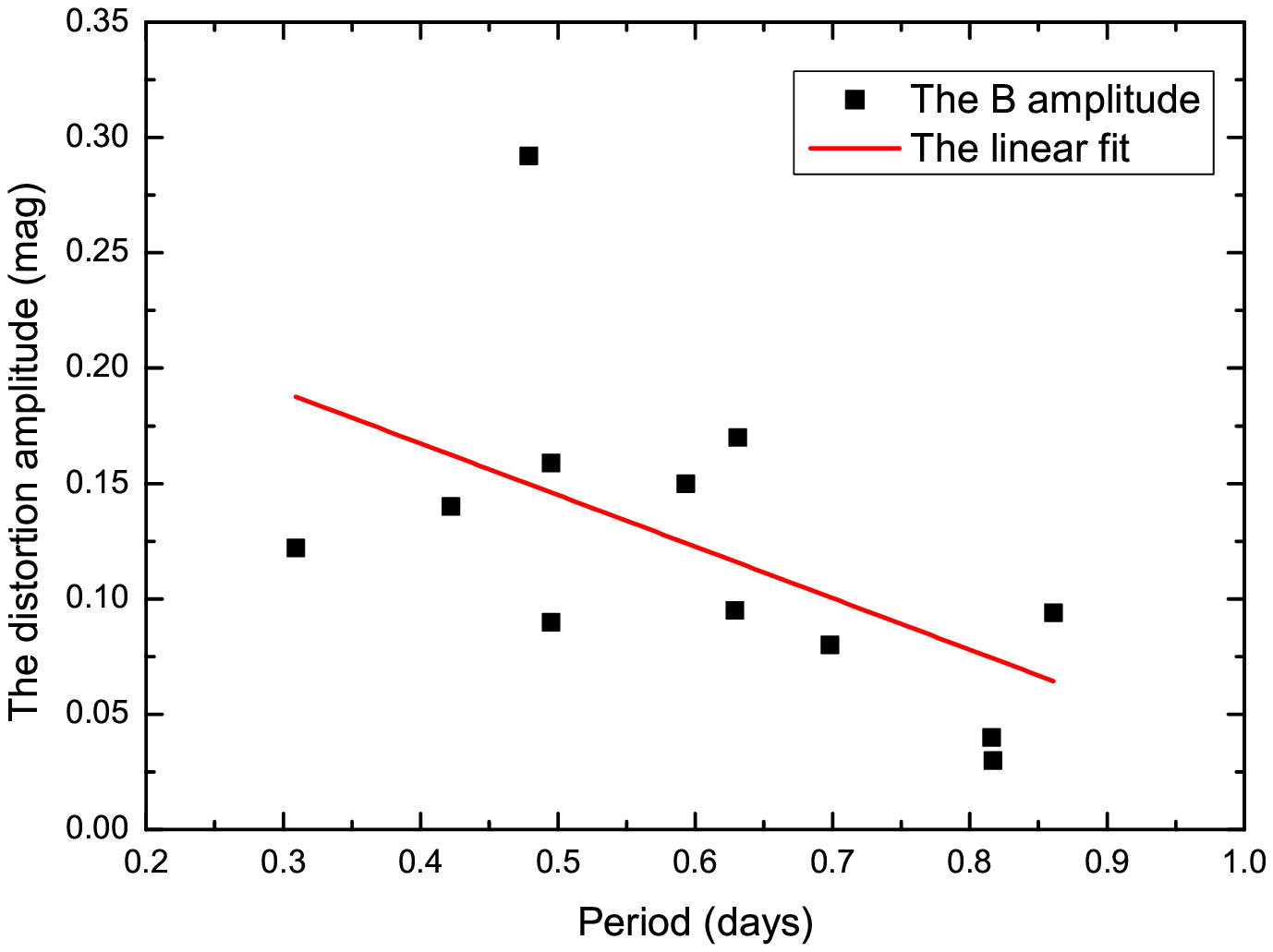}
\includegraphics[angle=0,scale=0.405]{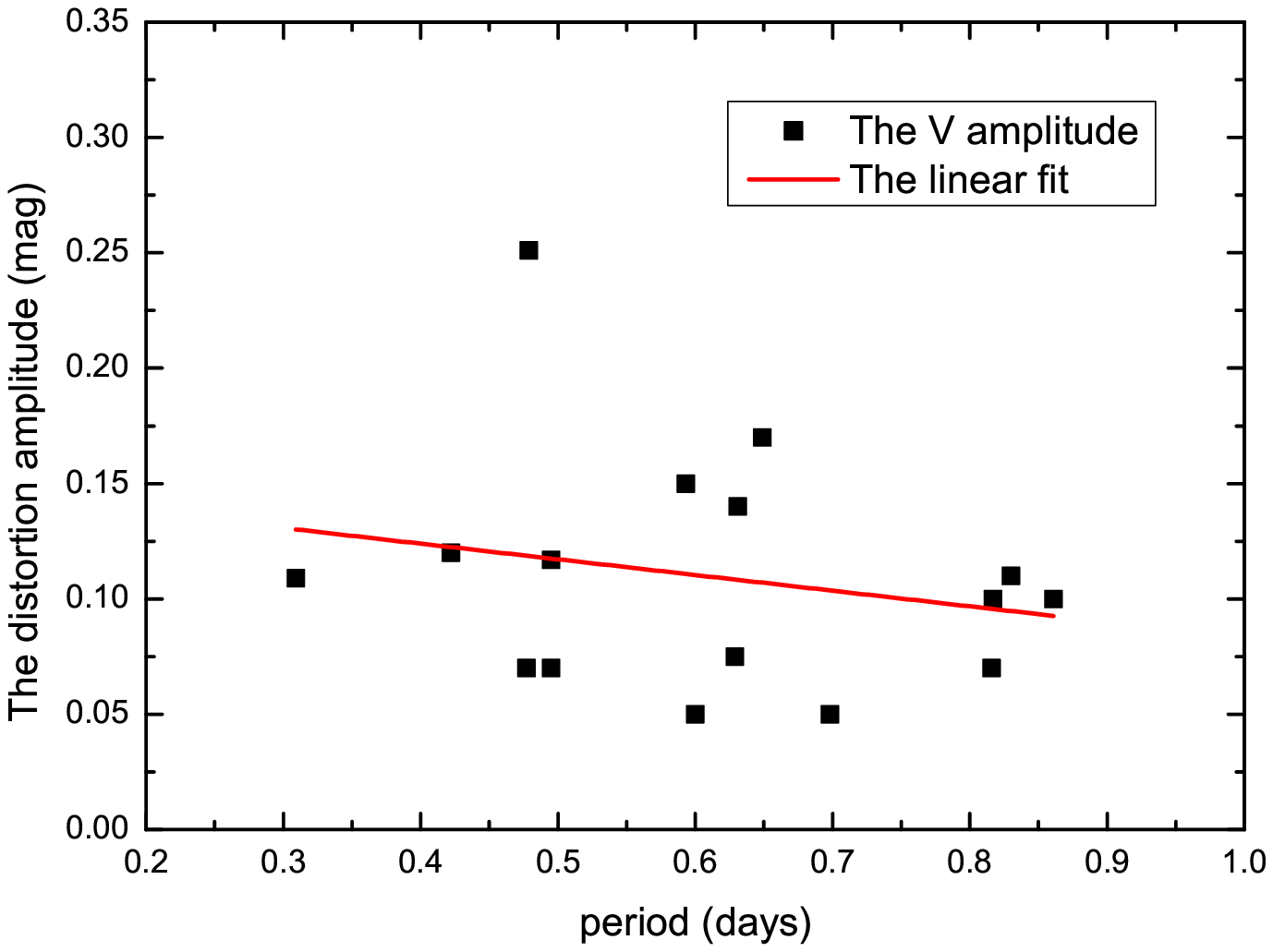}
\caption{The relations between the orbital period and photometric
distortion amplitudes in the B and V bands. }\label{fig:1}
\end{center}
\end{figure*}

\normalem
\begin{acknowledgements}
The authors would like to thank Fang X. S., for their kind helps,
and the observing assists of 85cm telescope of xinglong station for
their help during our observations. We also thank Profs. Zhou X.,
Jiang X. J., and Zhao Y. H., for allocation of observing time and
their kind helps during my visiting to NAOC. This work is partly
supported by GuiZhou University under grant No. 2008036, GuiZhou
natural Science Foundation 20092263 and supported partly by the
Joint Fund of Astronomy of the National Natural Science Foundation
of China (NSFC) grant No. 10978010.
\end{acknowledgements}

\appendix                  



\label{lastpage}

\end{document}